%% file: thors.tex
\documentclass{sigplanconf}

\usepackage{amsmath}
\usepackage{url}
\usepackage[usenames]{color}
\usepackage{tikz}
\usetikzlibrary{shapes,arrows,positioning,fit,matrix}
\usepackage{paralist}

\begin{document}

\conferenceinfo{ACM SIGPLAN Workshop on ML}{18th Sep 2011, Tokyo.} 
\copyrightyear{2011} 
\copyrightdata{[to be supplied]} 

\titlebanner{banner above paper title}        
\preprintfooter{short description of paper}   

\title{Verifying Liveness Properties of ML Programs}
\subtitle{ACM SIGPLAN Workshop on ML, 18th September 2011, Tokyo}

\authorinfo{M M Lester\and R P Neatherway\and C-H L Ong\and S J Ramsay}
           {Department of Computer Science, University of Oxford}

\maketitle

\input{abstract}

%
%

\input{introduction}

\input{algorithm}

\input{examples}

\input{future}

%

%


\bibliographystyle{abbrvnat}


\bibliography{thors}
%
%

\end{document}

%% file: abstract.tex
\begin{abstract}
Higher-order recursion schemes are a higher-order analogue of Boolean
Programs; they form a natural class of abstractions for functional
programs.  We present a new, efficient algorithm for checking CTL
properties of the trees generated by higher-order recursion schemes,
which is an extension of Kobayashi's intersection type-based model
checking technique.  We show that an implementation of this algorithm,
\textsc{Thors}, performs well on a number of small examples and we demonstrate
how it can be used to verify liveness properties of OCaml programs.
Example properties include statements such as ``all opened sockets are
eventually closed'' and ``the lock is held until the file is closed''.
\end{abstract}

%% file: introduction.tex
\section{Introduction}

Higher-Order Recursion Schemes (HORS) are a kind of higher-order tree grammar
for generating a (potentially infinite) tree. They are in essence closed, ground-type terms of the simply-typed lambda calculus with recursion and uninterpreted first-order constants. Because to the close relationship between the lambda-calculus and functional programming languages, HORS are a natural model of computation for functional programs. They provide, in particular, an extremely accurate account of higher-order
functions.  Moreover, HORS are well-suited to the purpose of verification
since they have a decidable mu-calculus model checking problem.  That is,
given a mu-calculus property $\phi$ and a HORS $\mathcal{G}$, the problem of
whether the tree generated by $\mathcal{G}$ satisfies $\phi$ can
be solved effectively, albeit with a rather challenging worst-case
time complexity: $n$-EXPTIME where $n$ is the largest order of any function in
$\mathcal{G}$~\cite{DBLP:conf/lics/Ong06}.

Following Kobayashi \cite{DBLP:conf/popl/Kobayashi09}, we aim to verify
properties of a given functional program by first constructing a HORS
$\mathcal{G}$ which generates the (possibly infinite) computation tree of
the program---i.e.~a tree whose paths represent runs of the
program that are labelled by observations of interest---and then model
checking $\mathcal{G}$.  Kobayashi restricted his attention to checking only
safety properties, but even in this more constrained setting the
model-checking problem is complete for $(n-1)$-EXPTIME.  However, in an
attempt to perform well outside of the worst-case, a follow-up paper
\cite{DBLP:conf/ppdp/Kobayashi09} presented an algorithm based on partial
evaluation and heuristic search which was shown to work remarkably well in
practice.

We have extended this approach to the verification of properties expressible
in the Alternation Free Mu-Calculus (AFMC), thus allowing for the
specification of \emph{both} safety \emph{and} liveness properties.  In
particular, this allows for the verification of every property expressible in the
Computation Tree Logic (CTL).  Our algorithm employs techniques similar to
those introduced by Kobayashi~\cite{DBLP:conf/ppdp/Kobayashi09} and, in addition,
comprises a weak B\"uchi game solver which has been heavily optimised for
our particular domain.

We have built an implementation of our algorithm, \textsc{Thors} (Types
for Higher Order Recursion Schemes), written in OCaml.  We have used
\textsc{Thors} to verify safety and liveness properties of a number of
interesting OCaml programs; the performance in our initial experiments has
been promising.  \textsc{Thors} can be tested through a web interface at
\url{https://mjolnir.comlab.ox.ac.uk/thors/}. A complete account is available in a technical report~\cite{tech:thors}.

%% file: algorithm.tex
\section{Algorithm}

\begin{figure}
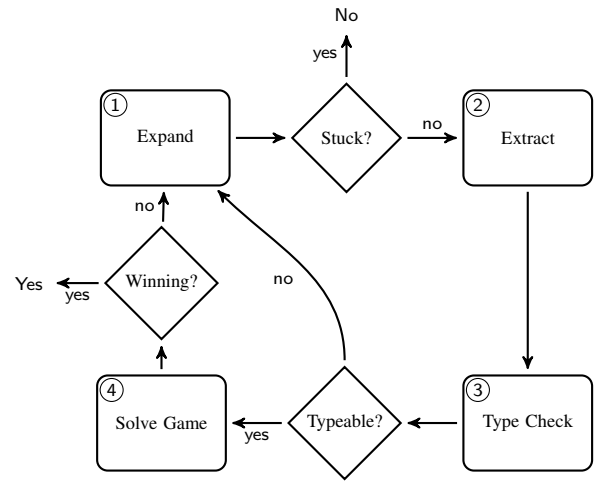

\include{alg_diagram}
\caption{Algorithm in outline}
\label{fig:algorithm}
\end{figure}

The core algorithm of our tool takes as input a HORS $\mathcal{G}$
and an Alternating Weak Tree automaton $\mathcal{A}$.
Typically, $\mathcal{G}$ will be an abstraction of the
functional program under consideration, generated from its function
definitions.
Meanwhile, $\mathcal{A}$ encodes the temporal logic property $\phi$ to be checked.
The algorithm decides whether $\mathcal{A}$ has an accepting run over
$\mathcal{G}$.

Our algorithm is based around an intersection type system, which is similar
to that of Kobayashi and Ong~\cite{DBLP:conf/lics/KobayashiO09}. Atomic types in the system are states of
$\mathcal{A}$. $\mathcal{G}$ is typeable in the system if and only if
$\mathcal{A}$ accepts the tree that $\mathcal{G}$ generates.
Typability depends not only on every function in $\mathcal{G}$ having a
valid typing, but also on there being a winning strategy in a certain parity
game constructed from the types and type environments used in this typing.
The size of the full parity game usually prohibits its explicit
construction. However, by forbidding weakening in our type system, we have
discovered that it is possible to consider only a small, relevant fragment
of the game, making its construction tractable in many cases.

The structure of the algorithm is shown in Figure 1.
In Stage 1, $\mathcal{G}$ is partially evaluated to obtain information about
the behaviour of the program.
If it is immediately apparent that a trace of the program violates the
property, then the algorithm terminates and reports that the problem is a No
instance.
Otherwise, the algorithm proceeds to Stage 2, where heuristics are used to
select candidate types for functions of $\mathcal{G}$ on the basis of the
partial evaluation.

Stage 3 type-checks the functions of $\mathcal{G}$ using these types.
If the candidate types are not self-consistent, types are discarded
until either a consistent set is found or all types have been discarded.
In the latter case, the algorithm returns to Stage 1 to evaluate the program
further and obtain more candiate types.
In the former case, the valid typing witnesses the
existence of a run tree of $\mathcal{A}$ over $\mathcal{G}$, but does not
determine whether this is accepting; intuitively, it checks safety, but
not liveness.

Thus the algorithm proceeds to Stage 4, which uses the types and the
corresponding type environments to construct a weak B\"uchi game (a kind of
parity game) for which the existence of a winning
strategy indicates that the run tree is accepting.
If a winning strategy exists, then the algorithm terminates and the problem
is a Yes instance. Otherwise, the algorithm returns to Stage 1 to evaluate
the program further in an attempt to find other run trees.

If $\mathcal{A}$ is deterministic, any valid run tree is unique, so instead
of looping back to Stage 1 from Stage 4, the algorithm can terminate and
return No. If $\mathcal{A}$ is non-deterministic and the problem is a
Yes instance, then termination is guaranteed. But on a No instance, the
algorithm may loop forever. However, we can solve this problem by running
a second copy of the algorithm in parallel on the complement automaton
$\overline{\mathcal{A}}$ and negating its result if it terminates first.

%% file: alg_diagram.tex
\tikzstyle{decision} = [diamond, draw,  
    text width=1cm, text badly centered, node distance=3cm, inner sep=2pt, font=\scriptsize, thick]
\tikzstyle{block} = [rectangle, draw, 
    text width=1.5cm, text centered, rounded corners, minimum height=4em, thick, font=\scriptsize]
\tikzstyle{line} = [style =-stealth', thick, shorten >= 2pt, shorten <= 2pt]

\begin{tikzpicture}[node distance=2cm, auto, font=\scriptsize\sf, >=angle 90]
\node [block] (expand) {Expand};
\node [circle, draw, above left = -0.32cm and -0.32cm of expand, inner sep=1pt] {1};
\node [decision, right = 0.8cm of expand] (stuck) {Stuck?};
\node [block, right = 0.8cm of stuck] (extract) {Extract};
\node [circle, draw, above left = -0.32cm and -0.32cm of extract, inner sep=1pt] {2};
\node [block, below = 2.5cm of extract] (check) {Type Check};
\node [circle, draw, above left = -0.32cm and -0.32cm of check, inner sep=1pt] {3};
\node [decision, left = 0.8cm of check] (typable) {Typeable?};
\node [block, left = 0.8cm of typable] (solve) {Solve Game};
\node [circle, draw, above left = -0.32cm and -0.32cm of solve, inner sep=1pt] {4};
\node [decision, above = 0.45cm of solve] (winning) {Winning?};
\node [left = 0.7cm of winning] (yes) {Yes};
\node [above = 0.7cm of stuck] (no) {No};
\node [left = 0.7cm of extract,draw=none] (fake) {};
\path [line] (expand) edge (stuck);
\path [line] (stuck) edge node {no} (extract);
\path [line] (extract) edge (check);
\path [line] (check) edge (typable);
\path [line] (typable) edge node {yes} (solve);
\path [line] (solve) edge (winning);
\path [line] (winning) edge node {no} (expand);
\path [line] (winning) edge node {yes} (yes);
\path [line] (stuck) edge node {yes} (no);
\path [line] (typable) edge[out=90, in=315] node {no} (expand);
\end{tikzpicture}

%% file: examples.tex
\section{Examples}

We discuss two examples constructed from ML programs. The full technique translates a Resource Usage Language~\cite{IgarashiK05} program directly to a HORS, using a bisimulation to prove correctness. Techniques for abstraction from ML are not covered in this paper. The translation to HORS uses a CPS transform to \begin{inparaenum}[(i)]\item{preserve ML call-by-value semantics in call-by-name HORS, and} \item{generate a computation tree of resource accesses}\end{inparaenum}.

\subsection{Intercept}

For this example, we take a network-oriented OCaml program. This program reads an arbitrary amount of data from a network socket into a queue and then forwards the data to another socket. The full program can be found online~\cite{www:intercept}; an abstracted form in ML-like syntax follows:

{\footnotesize\begin{verbatim}
let rec g y n = for i in 1 to n do write(y); done; close(y)
let rec f x y n = if b then read(x); f(x,y,n+1)
                       else close(x); g(y,n)
let t = open_out "socket2"
let s = open_in "socket1" in f(s,t,0)
\end{verbatim}}

For this program it would be useful to confirm that if the ``in'' socket stops transmitting data then the ``out'' socket is eventually closed
($AG \, {\it close}_{in} \Rightarrow AF \, {\it close}_{out}$). In order to distingush between these two resources, the alphabet in the image of the translation includes duplicate access primitives.

{\scriptsize
\[\begin{array}{rll}
S                & \rightarrow & {\it Newr} \,  {\it C1}\\
{\it C1} \, x          & \rightarrow & {\it Neww} \, ({\it C2} \, x) \\
{\it C2} \, x \, y     & \rightarrow & F \, x \, y \, {\it Zero} \, {\it end} \\
F  \, x \, y \, n \, k  & \rightarrow & {\it br} \, ({\it Read} \, x \, (F \, x \, y \, ({\it Succ} \, n) \, k)) \\
 & & \; \; \; \, ({\it Closer} \, x \, (G \, y \, n \, k) \\
G \, y \, n \, k & \rightarrow & n \, ({\it Write} \, y) \, ({\it Closew} \, y \, k) \\
I \, x \, y      & \rightarrow & x \, y \\
K \, x \, y      & \rightarrow & y \\
{\it Newr} \, k        & \rightarrow & {\it newr} \, (k \, I) \\
{\it Neww} \, k        & \rightarrow & {\it neww} \, (k \, I) \\
{\it Closer} \, x \, k & \rightarrow & x \, {\it closer} \, k\\
\end{array}\]
\[\begin{array}{rll}
{\it Closew} \, x \, k & \rightarrow & x \, {\it closew} \, k\\
{\it Read} \, x \, k & \rightarrow & x \, {\it read} \, k\\
{\it Write} \, x \, k & \rightarrow & x \, {\it write} \, k\\
{\it Zero} \, f \, x & \rightarrow & x\\
{\it Succ} \, n \, f \, x & \rightarrow & f \, (n \, f \, x)\\
\end{array}\]
}

\textsc{Thors} verifies that this HORS satisfies the property in 35ms. The scheme is order 4, while the property automaton has 2 states and the parity game has 31 nodes.

\subsection{Unbounded file access}

Our second example analyses a file with an unbounded number of file accesses. The program reads from a file for an unspecified length of time, before closing it and opening another.

{\footnotesize\begin{verbatim}
let rec g x = if b then close(x); g(open_in n)
                    else read(x); g(x) in
let s = open_in "foo" in g(s)
\end{verbatim}}

For this program, we wish to ensure that every opening of a file is followed by a finite number of reads and a close. Note that the program itself need not terminate. In CTL, we represent this with the property $AG \, ({\it newr} \Rightarrow AXA ({\it read} \ U  {\it close}))$. A translated version of the program is:

{\scriptsize
\[\begin{array}{rll}
S     & \rightarrow & {\it Newr} \, (G \, {\it end})\\
G \, k \, x & \rightarrow & {\it br} \, ({\it Close} \, x \, ({\it Newr} \, (G \, {\it end}))) \, ({\it Read} \, x \, (G \, k \, x))\\
I \, x \, y & \rightarrow & x \, y\\
K \, x \, y & \rightarrow & y\\
{\it Newr} \, k & \rightarrow & {\it brnew} \, ({\it newr} \, (k \, I)) \, (k \, K)\\
{\it Close} \, x \, k & \rightarrow & x \, {\it close} \, k\\
{\it Read} \, x \, k & \rightarrow & x \, {\it read} \, k\\
\end{array}\]
}

For the program to meet its specification we must, as is common when verifying liveness properties, impose a fairness constraint. Here we exclude any path containing an infinite sequence of reads, modelling an environment for our program that does not include files of infinite length.

\textsc{Thors} verifies that this HORS satisfies the property in 1ms. The scheme is order 4, while the property automaton has 3 states and the parity game has 17 nodes.

%% file: future.tex
\section{Future Work}

Although \textsc{Thors} performs well on many examples,
developing better heuristics would increase the range of
programs and properties we can practically verify.
Furthermore, the abstraction from OCaml programs to HORS is currently
manual; it would need to be automated in a practical verification tool.